\documentclass[UTF8,twocolumn]{aastex701}
\usepackage{ulem}
\usepackage{enumitem}
 \usepackage{amsmath}
 \usepackage{amssymb}

\newcommand{\bee}{\begin {equation}}
\newcommand{\ene}{\end {equation}}
\newcommand{\bqa}{\begin {eqnarray}}
\newcommand{\eqa}{\end {eqnarray}}
\usepackage{physics}
\usepackage{float}
\usepackage{natbib}
\received{\today}
\revised{\today}
\accepted{******}

\newcommand{\ie}{i.e.}  
\newcommand{\eg}{e.g.}
 
\newcommand{\etc}{etc.}

\shorttitle{Solar VIPA}
\shortauthors{Wang et al.}
\usepackage{graphicx}

\begin{document}

\title{A Compact, Ultra-High Resolution VIPA Spectrograph for Solar Spectroscopic Observations: Astrocomb Characterization and First Light}

\correspondingauthor{Xiaoming Zhu}
\email{xmzhu@niaot.ac.cn}

\author{Yutao Wang}
\email{ytwang@niaot.ac.cn}
\affiliation{Laboratory of Solar and Space Instruments, Nanjing Institute of Astronomical Optics $\&$ Technology, Chinese Academy of Sciences, Nanjing 210042, China}
\affiliation{CAS Key Laboratory of Astronomical Optics $\&$ Technology, Nanjing Institute of Astronomical Optics $\&$ Technology, Nanjing 210042, China}
\affiliation{University of Chinese Academy of Sciences, Beijing 100049, China}

\author[0000-0003-1088-1457]{Xiaoming Zhu}
\email{xmzhu@niaot.ac.cn}
\affiliation{Laboratory of Solar and Space Instruments, Nanjing Institute of Astronomical Optics $\&$ Technology, Chinese Academy of Sciences, Nanjing 210042, China}
\affiliation{CAS Key Laboratory of Astronomical Optics $\&$ Technology, Nanjing Institute of Astronomical Optics $\&$ Technology, Nanjing 210042, China}

\author[0000-0003-2891-6267]{Xiaoli Yan}
\email{yanxl@ynao.ac.cn}
\affiliation{Yunnan Observatories, Chinese Academy of Sciences, Kunming 650216, China}
\affiliation{Yunnan Key Laboratory of Solar Physics and Space Science, Kunming, 650216, China}

\author[0000-0002-7289-642X]{P. F.  Chen}
\email{chenpf@nju.edu.cn}
\affiliation{School of Astronomy and Space Science, Nanjing University, Nanjing 210023, China}
\affiliation{Key Laboratory of Modern Astronomy and Astrophysics (Nanjing University), Ministry of Education, Nanjing 210023, China}

\author{Huiqi Ye}
\email{hqye@niaot.ac.cn}
\affiliation{CAS Key Laboratory of Astronomical Optics $\&$ Technology, Nanjing Institute of Astronomical Optics $\&$ Technology, Nanjing 210042, China}
\affiliation{Laboratory of astronomical spectra and high-resolution imaging, Nanjing Institute of Astronomical Optics $\&$ Technology, Chinese Academy of Sciences, Nanjing 210042, China}
\author{Dong Xiao}
\email{dxiao@niaot.ac.cn}
\affiliation{CAS Key Laboratory of Astronomical Optics $\&$ Technology, Nanjing Institute of Astronomical Optics $\&$ Technology, Nanjing 210042, China}
\affiliation{University of Chinese Academy of Sciences, Beijing 100049, China}
\affiliation{Laboratory of astronomical spectra and high-resolution imaging, Nanjing Institute of Astronomical Optics $\&$ Technology, Chinese Academy of Sciences, Nanjing 210042, China}

\affiliation{University of Chinese Academy of Sciences, Nanjing 211135, China}

\author[0000-0002-1899-3384]{Jinping He}
\email{jphe@niaot.ac.cn}
\affiliation{Laboratory of Solar and Space Instruments, Nanjing Institute of Astronomical Optics $\&$ Technology, Chinese Academy of Sciences, Nanjing 210042, China}
\affiliation{CAS Key Laboratory of Astronomical Optics $\&$ Technology, Nanjing Institute of Astronomical Optics $\&$ Technology, Nanjing 210042, China}
\affiliation{University of Chinese Academy of Sciences, Beijing 100049, China}
\affiliation{University of Chinese Academy of Sciences, Nanjing 211135, China}

\correspondingauthor{Jinping He}
\email{jphe@niaot.ac.cn}

\begin{abstract}

We present a compact, high spectral resolution prototype spectrograph based on a Virtually Imaged Phased Array (VIPA), which is designed for solar spectral observations. This fiber-fed instrument has a size of only 53 $\times$ 20 $\times$ 18 cm$^3$. Wavelength calibration using an astrocomb ($f_{\text{rep}}=25$ GHz) established an operational bandpass of 592.76--657.07 nm and revealed an asymmetric instrumental profile. A Fano-Lorentz product function provides a significantly better fit to this profile than a Gaussian. The measured spectral resolution ranges between 290,000 and 340,000 across the band. Initial on-sky validation at the New Vacuum Solar Telescope (NVST, Yunnan Observatories) successfully demonstrated the prototype's capabilities: clear detection of solar five-minute oscillations ($\pm 300 \, \text{m s}^{-1}$) in the \ion {Fe}{1} 6280.57 \AA~ line, resolution of magnetic broadening in sunspots using the \ion{Fe}{1} 6173.34  \AA~ line, and the first ground-based definitive identification of the faint \ion{Si}{1} 6560.57 \AA~ line within the H$\alpha$ band. 
These results validate the VIPA as a promising platform for high spectral resolution solar spectroscopy. Its compact design and performance directly support future applications in multi-object solar studies, high spectral resolution observations for time-domain astronomy, including exoplanet detection, and potential space-borne instrumentation.

\end{abstract}



\section{Introduction} 

High-resolution spectroscopy remains an indispensable tool in solar physics, providing critical diagnostics for characterizing magnetohydrodynamic processes across the solar atmosphere. 
For instance, the \ion{Fe}{1} 6173.34 \AA~can resolve photospheric magnetic fields down to $\sim$ 10 G in sunspot umbrae via Zeeman splitting \citep[\eg,][]{smitha2023non}, while the \ion{He}{1} 10830 \AA~line provides essential information to study the fibril dynamics and solar flares in chromosphere \citep[\eg,][]{wang2021high,du2008properties}. 
For velocity fields, Doppler shifts measured at sub-km s$^{-1}$ accuracy reveal convective blueshifts \citep[\eg,][]{lohner2018convective}, p-mode oscillations \citep[\eg,][]{lemke2016gottingen}, and shock waves in spicules \citep[\eg,][]{de2007chromospheric}. 
Additionally, spectral line asymmetries observed from center-to-limb variations help reveal radiative transfer effects within the solar atmosphere \citep[\eg,][]{stenflo2015fts}. 
Crucially, these measurements rely on ultra-stable wavelength calibration, \ie astrocombs, which achieve precision better than 10 cm s$^{-1}$ \citep[\eg,][]{doerr2012laser}, enabling the detection of mass flows even in the quiescent state \citep[\eg,][]{liu2013horizontal}. 

The aforementioned observations all require a spectral resolution $\mathcal{R} \ge 300,000$. Conventional \'{e}chelle spectrographs can achieve such resolutions but require large-scale optical setups and stringent thermal/vibrational stability  \citep[\eg,][]{pepe2000harps,conconi2013espresso}, restricting their application in solar diagnostics. 
The Virtually Imaged Phased Array (VIPA) \citep[\eg,][]{shirasaki1996large} provides a promising alternative for high-resolution spectroscopy. Its ultra-high angular dispersion facilitates a highly compact spectrograph architecture \citep[\eg,][]{bourdarot2017nano,zhu2020vipa}. Furthermore, the spectral resolution achieved with a VIPA spectrograph exhibits minimal dependence on the input slit width/fiber core diameter \citep[\eg,][]{zhu2023dispersion}. Although VIPA has found widespread application in laser spectroscopy, precision ranging, and Brillouin spectroscopy, \etc , its application in astronomical fields, especially for solar observations, is still nascent.

In this work, we present the development, astrocomb-based characterization, and first on-sky validation of a high spectral resolution, multimode-fiber-fed VIPA spectrograph for solar observations, demonstrating its feasibility for solar physics research.
The paper is structured as follows.
Section 2 details the optical design and mechanical construction of the compact VIPA spectrograph. Section 3 presents the wavelength calibration using an astrocomb and characterizes the asymmetric point spread function (PSF). Section 4 describes the first-light solar observations and presents the detection of previously obscured silicon lines near the H$\alpha$ band. Finally, Section 5 provides a comprehensive discussion of the instrument’s performance, scientific implications, and future prospects.

\section{Optical Design and Laboratory Integration} \label{sec:design}
\begin{figure}[h]
\centering\includegraphics[width=8.4cm]{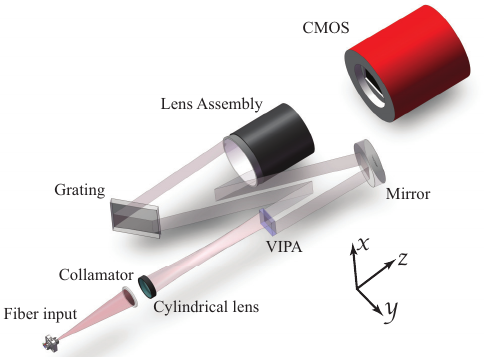}
\caption{Optical layout of the compact VIPA spectrograph. The instrument accepts both single-mode and multimode fiber inputs and integrates all components into a unit measuring $53 \times 20 \times 18\,\text{cm}^3$.}
\label{fsetup}
\end{figure}

Figure \ref{fsetup} shows the design layout of the VIPA spectrograph. The instrument accepts fiber-optic input and supports both single-mode fibers (SMFs) and multimode fibers (MMFs) with a core diameter $\varnothing\le$ 25 $\mu$m and a numerical aperture (NA) of 0.1.
After entering the spectrograph, the optical signal is collimated first and then focused by a cylindrical lens before being coupled into the VIPA. 
The VIPA is a commercial one purchased from LightMachinery Inc with a volume of 22 $\times$ 24 $\times$ 1.68 mm$^3$ filled with fused silica and the working wavelength range of 600 -- 750 nm.
 The VIPA achieved a throughput of 90\% through the careful selection of focal lengths of the collimator ($f_s=100$ mm) and the cylindrical lens ($f_c=150$ mm) with a $\varnothing$25 $\mu$m MMF input while maintaining the compact design. 
 
The orthogonal dispersion element employs a reflective holographic grating, 2400 g/mm, 50 $\times$ 50 mm$^2$, {which helps to minimize the generation of stray light, despite its lower efficiency}. For broad-spectrum measurements, an imaging lens assembly ($f_i=178.58$ mm) is designed to minimize aberrations and a ZWO ASI6200MM CMOS camera is utilized as the detector.  
The CMOS has a 3.76 $\mu$m pixel pitch with 6388 pixels in the VIPA's  dispersion direction and 9576 pixels in the grating's.
All components are integrated into a compact unit with the dimensions of $53 \times  20 \times 18$ cm$^3$.

\section{Astrocomb-Calibrated Performance Characterization} \label{sec:cali}
The exceptional frequency precision and narrow linewidth of the astrocomb make it an ideal source for precise characterization of two key spectrograph properties: the wavelength scale and the instrumental profile. The astrocomb used to calibrate our VIPA spectrograph is a system from Menlo Systems GmbH installed on the High-Resolution Spectrograph (HRS) of the Chinese 2.16 m telescope at Xinglong Observatory. It operates at a repetition frequency ($f_{rep}$) of 25 GHz and a carrier-envelope offset (CEO) frequency of 6.27 GHz. This section details the calibration procedures performed with this astrocomb and presents the derived performance parameters. 
Furthermore, our spectrograph's compatibility with both SMF and MMF inputs allows for a direct comparison of its dispersion characteristics under the respective conditions.

\subsection{Characterization of the Instrument Profile}
\label{cip}
The array of multiple beams exiting the rear surface of the VIPA is characterized by four primary features: 
\begin{itemize}[label={\fontsize{4}{4}\selectfont\textbullet}]
\item each beam is divergent with an identical numerical aperture (NA);
\item there is a constant optical path difference between successive beams;
\item they propagate mutually parallel to one another;
\item their amplitude exhibits a non-linear decay along the dispersion direction.
\end{itemize}
These distinctive characteristics, which fundamentally differ from those of conventional gratings, result in an instrumental profile (IP) for the VIPA spectrograph with unique properties. In the present setup, the separations between astrocomb teeth are larger than 15 full-width at half maximum (FWHMs), enabling a detailed investigation of the characteristics inherent to the VIPA spectrograph's instrumental profile.

\begin{figure}[h]
\centering\includegraphics[width=8.4cm]{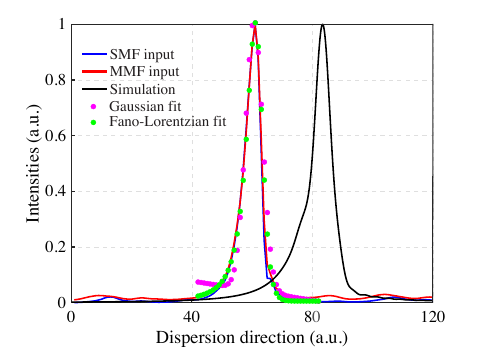}
\caption{Measured and simulated instrumental profiles (IPs) of the VIPA spectrograph for both single-mode (SMF) and multimode fiber (MMF) inputs. The pronounced asymmetry, well-fitted by a Fano--Lorentz product function, is intrinsic to the VIPA's operation.}
\label{psf}
\end{figure}

In Figure \ref{psf}, the averaged IPs obtained with SMF and MMF input, together with simulation results are presented. {The simulation accounts for the four aforementioned characteristics of the output beam and incorporates the impact of non‑ideal lenses in a practical setup by modelling the lenses with realistic thickness, finite aperture, and the associated aberrations, thus allowing their effects to be evaluated and quantified.} Data for both experiments and simulations were acquired near the optical axis, and both sets of results consistently demonstrate the asymmetric nature of the VIPA's IP.

The asymmetry of the VIPA spectrograph's IP originates from both its fundamental operating principles and practical experimental factors.
The dispersion of a VIPA arises from multi-beam interference. For monochromatic light at a wavelength $\lambda$, the signal recorded on the spectrograph detector corresponds to the resonance maxima formed by the interference in the focal plane of the imaging lens. When a resonance maximum coincides with the lens focal point, it also acts as the lens's diffraction spot. Imperfections in spherical optics and the aperture constraints in a real optical system cause these spots to spread spatially, with the spreading depending on off‑axis distance.

Moreover, the energy distribution of the multiple beams along the dispersion direction ($\hat x$ in Figure \ref{fsetup}) follows an approximately exponential decay, $r^n\approx\exp[-(1-r) n]$, where $n\in N$ and $r > 0.9$ is the partial reflectivity of the VIPA rear side. This distribution is inherently asymmetric with respect to the optical axis, and its energy centroid is offset. Consequently, the spatially broadened resonance maxima exhibit a monotonic intensity profile in the $\hat x$-direction, directly leading to the observed asymmetry in the instrumental profile. This asymmetry is therefore intrinsic to the VIPA’s light-propagation mechanism and is largely unavoidable.

In practice, further factors can modify or exacerbate the observed asymmetry. These include:
\begin{itemize}[label={\fontsize{4}{4}\selectfont\textbullet}]
\item Non‑ideal lens optics (e.g., aberrations),

\item Slight optical misalignments,

\item Etalon surface imperfections, such as non‑parallelism between the VIPA’s front and rear surfaces.
\end{itemize}
These experimental imperfections collectively influence the measured interference pattern and often amplify the underlying asymmetric line shape.

{Under ideal conditions with perfect lenses acting as Fourier transformers, the interference maxima of a VIPA exhibit a symmetric Lorentzian line shape \citep[\eg,][]{xiao2004dispersion,zhu2020vipa,zhu2023dispersion}. In a practical setup, however, the combination of lens imperfections (finite thickness, finite aperture, and aberrations) and the intensity attenuation along the rear surface of the VIPA modulates the line shape and introduces an unavoidable asymmetry, which necessitates a modified model.} To account for this, we employ a Fano function  as a modulator of the Lorentzian profile, leading to the Fano-Lorentz product function:
\bee
f_{FL}=a_1\frac{(\frac{x-\mu}{\gamma}-q)^2}{\Big[(\frac{x-\mu}{\gamma})^2+1\Big]^2}+a_2.
\ene
In our case, $q$ is close to 1. Fortunately, the line center of this fitting function is analytically tractable as
\bee
x_c=\mu+\gamma q-\gamma\sqrt{1+q^2}.
\ene

A total of 14,688 comb teeth was fitted using the Levenberg–Marquardt algorithm. For comparison, a Gaussian function with an additive linear polynomial was also employed \citep[\eg,][]{hao2018calibration,zhu2020vipa}. As shown in Figure \ref{psf}, the Fano–Lorentz product provides a markedly better description of the VIPA spectrograph’s IP, particularly in reproducing the asymmetric wings of the profile.

Owing to the analytical complexity of the line center position, which precludes conventional error propagation via the Jacobian matrix, we employed a Monte Carlo approach to evaluate its uncertainty. This involved generating 10,000 synthetic spectra by adding Poisson-distributed noise to the best Fano-Lorentz production fit, with each spectrum subsequently refitted using the same model. The standard deviation of the resulting 10,000 line center values was adopted as the final uncertainty in the Fano-Lorentz production line center. Our analysis confirmed that this ensemble size was sufficient, yielding a convergence error below 0.05\% and thus meeting our precision requirements.

For comparison, the same Monte Carlo procedure was applied to evaluate the line center uncertainty using the polynomial-added Gaussian fitting function. The resulting positional noise levels (PNL) were 1.80 m s$^{-1}$ (from the Jacobian matrix with Gaussian), 1.94 m s$^{-1}$ (Monte Carlo approach with Gaussian), and 2.64 m s$^{-1}$ (Fano–Lorentz). This indicates that while Fano-Lorentz production model more accurately describes the overall line shape of the VIPA spectrograph's IP, the Gaussian fit offers superior precision in determining the line center position. A trend is also observed in tests with pure Lorentzian, Fano, and hyperbolic secant functions, where the Gaussian consistently gave the smallest uncertainty. This intriguing phenomenon warrants further detailed investigation in future work. All subsequent data presented in this work are obtained from the Fano-Lorentz production model fits.

\subsection{Wavelength Calibration and Coverage}
The wavelength calibration was performed following the methods described by \citet{hao2018calibration} and \citet{zhu2020vipa}, which allows a precise wavelength-pixel solution. This process utilized the astrocomb in conjunction with a He-Ne laser. The well-known frequency of the He-Ne laser serves as a fixed reference to identify the absolute mode order $N$ of the astrocomb teeth by applying the relation $f=f_{ceo}+N f_{rep}$. With the mode numbers securely identified across the entire detector, the operational wavelength range of the VIPA spectrograph is from 592.76 nm to 657.07 nm. 

\begin{figure}[h]
\centering\includegraphics[width=8.4cm]{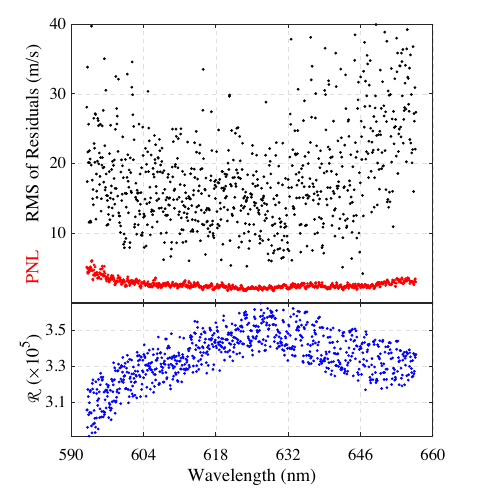}
\caption{Top panel: Absolute wavelength calibration accuracy (residual RMS) and the fundamental photon-noise limit (PNL) across the operational bandpass. Bottom panel: Spectral resolution ($\mathcal{R}$) derived from astrocomb line fitting, showing a non-monotonic variation with wavelength}
\label{resid}
\end{figure}

The calibrated range encompasses 816 distinct VIPA dispersion orders, with the free spectral range (FSR) of individual orders varying from 0.7076 \AA~to 0.8694 \AA~across the band. The residual RMS of the wavelength-pixel fit and the photon noise limit (PNL), evaluated throughout all orders, are presented in Figure \ref{resid}. 

Although the adoption of a Fano-Lorentz production model considerably improved the instrumental profile characterization and the resultant wavelength calibration over a Gaussian model, the achieved accuracy of 18.13 m s$^{-1}$ is nevertheless much worse than the fundamental photon noise limit of 2.64 m s$^{-1}$, as shown in the top panel of Figure \ref{resid}. {This discrepancy is expected and arises from systematic errors beyond fundamental photon noise, as discussed below.}

{A primary factor is the mismatch between the 25 GHz astrocomb spacing and the VIPA spectrograph's spectral resolution. The resulting line separation is approximately 16.6 FWHMs, which is much larger than the typical 2.5--3 FWHMs spacing in astronomical spectrographs optimized for astrocomb calibration. In our setup, in each VIPA diffraction order only 18 comb teeth were involved in the wavelength‑pixel polynomial fit. For comparison, when the same astrocomb was used on a conventional \'{e}chelle spectrograph (\eg, HRS on the 2.16 m telescope, \citealp{hao2018calibration}), each order contained 150--300 comb teeth, allowing a well‑constrained fit. In our VIPA spectrograph, such sparse sampling severely under‑constrains the 5th--order polynomial, and the resulting interpolation/extrapolation uncertainty directly dominates the calibration accuracy. Therefore, this sparse sampling is the primary systematic error limiting the achieved accuracy.}

{It is noted that other systematic errors, such as optical aberrations, detector effects, and background subtraction residuals, are also present but are minor. Thus, the final calibration accuracy is dominated by the mismatch between the comb spacing and the VIPA’s dispersion characteristics, rather than by photon noise.}

\subsection{Spectral Resolution Measured by Comb Lines}
The astrocomb provides an ideal source for a direct and precise measurement of the spectrograph's spectral resolution, defined as $\mathcal{R} = \lambda / \Delta\lambda$, where $\Delta\lambda$ is the FWHM of the instrumental profile. We determined $\Delta\lambda$ by fitting the instrumental profile model to individual, unblended astrocomb lines. Given the comb's exceptional stability and negligible intrinsic width, the measured $\Delta\lambda$ faithfully represents the instrument's dispersion capabilities.

The resulting resolving power across the bandpass is presented in the bottom panel of Figure \ref{resid}. 
The theoretical resolving power of a VIPA is given by:
\bee
\mathcal{R} = \mathcal{F}  \frac{2 n_r t}{\lambda},
\ene
where $\mathcal{F}$ is the Finesse of the VIPA, $n_r$ the refractive index, and $t$ the VIPA thickness. The Finesse is primarily governed by the coating reflectivity. This relation predicts a monotonic decrease in resolution with increasing wavelength.

However, our measured data reveal a more complex behavior: $\mathcal{R}$ varies between approximately 290,000 and 340,000, exhibiting a non-monotonic trend with a peak near the center wavelength of 623.15 nm. This deviation from ideal behavior can be attributed to a combination of physical effects. First, the VIPA's dispersion characteristics and its high-reflectivity coating are optimized for a specific wavelength range. The degradation in resolution at the short-wavelength end (below $\sim$ 600 nm) is likely influenced by the declining performance of the VIPA's dielectric coating outside its nominal design bandpass (600 -- 750 nm). This can lead to a reduction in the effective finesse of the etalon, broadening the instrumental profile.

Second, more significantly, the optical system's off-axis aberrations play a crucial role. In our cross-dispersed configuration, different wavelengths within a single VIPA order impinge on the camera at different field angles. At the short-wavelength extremity of the spectrum, the beam is more heavily tilted relative to the optical axis of the imaging lens. This exacerbates aberrations such as astigmatism and coma, which spatially broaden the point spread function and consequently degrade the spectral resolution. Therefore, the observed non-monotonic trend in resolving power is attributed to the combined effects of wavelength-dependent coating performance and off-axis optical aberrations.

The above analysis underscores that the ultimate resolution of a practical VIPA spectrograph is not solely determined by the VIPA itself but is also a function of the ancillary optics. Future iterations of the instrument could achieve more uniform performance across a broad bandpass by incorporating custom-designed VIPA coatings and utilizing imaging optics with superior off-axis aberration correction.

A comparative analysis further indicates that the spectral resolution obtained with a multimode fiber input is 4.27\% lower than that achieved with a single-mode fiber. 
This finding, consistent with the results of \citet{zhu2023dispersion}, provides quantitative support for our earlier proposition regarding the inherent robustness of the VIPA spectrograph's resolution against variations in input conditions.\\

\section{First Light Observations with the NVST}

The first on-sky validation of the VIPA spectrograph was conducted at the 1-meter New Vacuum Solar Telescope (NVST, Yunnan Observatories) on 2024 March 16 and 17, with observations conducted from {02:00 to 08:00} UT each day. Observations were carried out at the F/45 focal station. The spectrograph coupling optics yielded a field of view of approximately $10\arcsec$ on the solar disk, corresponding to a linear scale of about 7250 km. Under these conditions, with a detector integration time of 1 second, the spectrograph achieved a signal-to-noise ratio greater than 145.

Our VIPA spectrograph simultaneously covers a spectral range of 592 -- 657 nm, resolving approximately 650 distinct solar spectral lines within a single exposure over a field of view of a few arcseconds on the solar disk. Although the astrocomb was not deployed for real-time calibration during these on-sky observations, the abundant telluric absorption lines from atmospheric water vapor and oxygen within the instrument's bandpass served as a reliable wavelength reference. This approach proved particularly effective given the observing conditions at the lakeside NVST site.

We successfully acquired a series of spectroscopic datasets during the on-sky validation. The instrument performance was verified through observations targeting different solar features and dynamical phenomena. The analysis of these datasets is presented in the following subsections.
\subsection{Solar Five-Minute Oscillations}
The solar five-minute oscillations, a well-known phenomenon, were clearly detected in the quiet-Sun regions within our field of view. The high spectral resolution and stability of the VIPA spectrograph enabled precise Doppler shift measurements across its entire spectral range. For this study, we selected the \ion{Fe}{1} 6280.57 \AA~line as our primary target. This choice was strategic, as this photospheric line is located near two telluric oxygen absorption lines at 6279.897 \AA~and 6280.394 \AA. This arrangement provided a stable, in-situ wavelength reference, which we used to accurately calibrate and remove the instrumental drift of the spectrograph.

\begin{figure}
\centering\includegraphics[width=8.4cm]{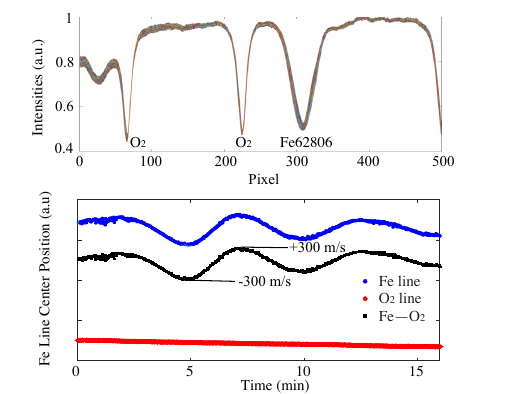}
\caption{Detection of solar five-minute oscillations. Top panel: Time series of 842 consecutive spectra around the \ion{Fe}{1} 6280.57 \AA\ line. Bottom panel: Derived Doppler velocity of the \ion{Fe}{1} line: the blue line shows the raw measured shift of the Fe line, the red line indicates the shift of the telluric O$_2$ line (used as a reference for spectrograph drift), and the black line represents the corrected Fe line shift after removal of the O$_2$ line drift, revealing oscillations with an amplitude of $\sim\pm300$\,m\,s$^{-1}$.}
\label{Osci5m}
\end{figure}

We conducted a continuous 16-minute observation of a quiet-Sun region near the solar disk center with a cadence of approximately 52 frames per minute. The resulting sequence of over 842 spectral images around the \ion{Fe}{1}  line is compiled in the top panel of Figure~\ref{Osci5m}. A direct comparison reveals that the instrumental drift derived from the two oxygen telluric lines is significantly smaller than the observed Doppler shift of the \ion{Fe}{1} line, confirming their efficacy as a stable wavelength reference. The processed Doppler velocity of the \ion{Fe}{1} line, plotted in the bottom panel of Figure~\ref{Osci5m}, exhibits clear periodic oscillations with an amplitude of approximately $\pm$300 m s$^{-1}$ and a period of about five minutes. This result is a clear signature of the solar five-minute oscillations.

\subsection{The \ion{Fe}{1} 6173 Line in Sunspot, Filament, and Quiet Region}
The \ion{Fe}{1} 6173.34 \AA~ line, a well-known photospheric line with high magnetic sensitivity (\textit{g}$_{\rm eff}=2.5$), serves as an excellent diagnostic tool for probing the thermal and magnetic structures of various solar structures.{We acquired the spectra of the VIPA diffraction order covering the range of 6172.586--6174.121 \AA, which includes the \ion{Fe}{1} 6173.34 \AA~line, through sequential observations of a sunspot umbra, a dark filament, and a quiet‑Sun region, facilitating a comparative analysis of their physical properties.}

\begin{figure}[h]
\centering\includegraphics[width=8.4cm]{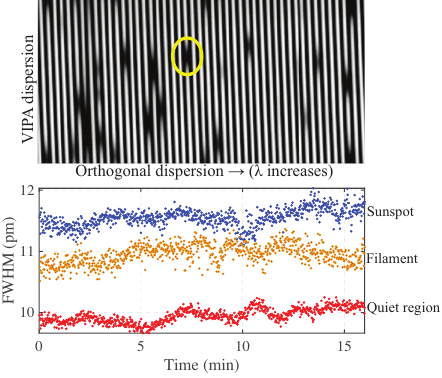}
\caption{Spectroscopic analysis of the \ion{Fe}{1} 6173.34\,\AA\ line in different solar features. Top panel: A partial raw spectral image, with the line position circled. Bottom panel: Temporal variation of the line's full width at half maximum (FWHM) in a sunspot umbra, a dark filament, and a quiet-Sun region, demonstrating magnetic and thermal broadening.}
\label{fe6173}
\end{figure}

The top panel of Figure~\ref{fe6173} displays the raw two-dimensional spectral image acquired by the VIPA spectrograph's detector, with the location of the \ion{Fe}{1} 6173.34 \AA~line indicated by the yellow circle. The bottom panel showcases the measured FWHM of the \ion{Fe}{1} line profiles extracted from the three distinct regions. A clear line broadening is evident in the sunspot region compared to the dark filament and quiet-Sun areas. The average line widths during the observation were measured to be 11.55, 10.93, and 9.95 pm for the sunspot, dark filament, and quiet-Sun region, respectively.

The significant broadening of the \ion{Fe}{1} 6173.34 \AA\ line within the sunspot umbra can be qualitatively attributed to two primary effects arising from the strong magnetic field and the altered thermal environment. First, the presence of a strong magnetic field induces Zeeman splitting, which effectively broadens the observed line profile \citep{unno1956line}. Second, the enhanced Doppler broadening due to increased turbulence and the modified thermodynamic state (e.g., lower temperature and higher microturbulence) within the sunspot also contribute to the overall line width \citep{solanki2003sunspots}. This combined effect results in the pronounced broadening observed in the sunspot spectrum.

Our VIPA spectrograph clearly resolved the broadening of the photospheric \ion{Fe}{1} line in the dark filament region compared to the quiet Sun. This subtle spectroscopic signature, successfully detected by our instrument, demonstrates its high sensitivity for diagnosing microscale physical processes responsible for line broadening. The observed line broadening is a known effect attributed to the presence of a magnetic flux rope hosting solar filaments \citep{lites2005magnetic}, which modifies the photospheric and chromospheric environment by enhancing turbulence and introducing complex plasma flows \citep{kuckein2012active}. A related debate is about the mechanism of coronal heating, \ie, waves or magnetic reconnection. To distinguish the competing mechanisms, precise measurements of line broadening of photospheric and coronal lines are strongly needed \citep{zirker1993coronal}. This capability to discern such fine spectral features in different solar features underscores the VIPA spectrograph's value as a powerful tool for high-resolution solar spectroscopic observations.

\subsection{Identification of a \ion{Si}{1} Line near the H$\alpha$ Band}

\begin{figure}
\centering\includegraphics[width=8.2cm]{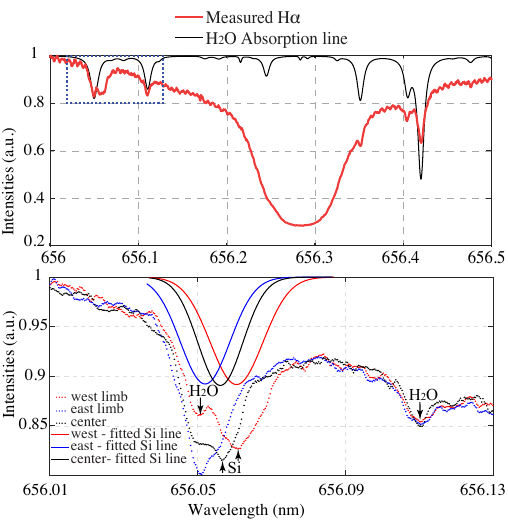}
\caption{First ground-based definitive detection of the faint photospheric \ion{Si}{1} 6560.57 \AA\ line within the H$\alpha$ band. Top panel: Full spectrum from 6560 to 6565 \AA, overplotted with telluric H$_2$O lines from the HITRAN database. Bottom panel: Detailed profiles at the solar disk center, east limb, and west limb, showing the Doppler-shifted \ion{Si}{1} line relative to stationary telluric features. The dotted lines are from the observations, and the solid lines are the Si I 6560.57 \AA\ spectral profiles from the west limb, disk center, and east limb extracted from the observations.}
\label{halpha}
\end{figure}

In the far blue wing of the H$\alpha$ spectral profile lies a silicon line, \ie, \ion{Si}{1} 6560.57 \AA\ \citep[\eg,][]{radziemski1965arc}. However, this line is blended with a telluric line, \ie, H$_2$O 6560.50 \AA. Therefore, the \ion{Si}{1} 6560.57 \AA\ line was frequently mis-identified as the telluric water line in the ground-based spectral observations. After the Chinese H$\alpha$ Solar Explorer (CHASE) satellite was launched in 2021, it revealed the \ion{Si}{1} 6560.57 \AA\ line clearly in the blue wing of the H$\alpha$ spectral profile \citep[\eg,][]{li2022chinese,qiu2022calibration,hong2022statistical}, as there is no contamination from the terrestrial atmosphere. 
A paramount and distinct result of this commissioning run, uniquely enabled by the instrument's combined high spectral resolution, was the first ground-based definitive detection of the faint \ion{Si}{1} 6560.57~\AA~line within the H$\alpha$ band, thanks to the high spectral resolution of our spectrometer. Figure~\ref{halpha} depicts this discovery. The top panel displays the full spectral range from 6560 to 6565~\AA~acquired by the VIPA spectrograph, with the corresponding telluric water vapor absorption lines from the HITRAN database overplotted for reference. The wavelength scale was absolutely calibrated using the two well-isolated telluric oxygen absorption lines at 6560.50 and 6561.10~\AA. The bottom panel provides a detailed view of the critical region between 6560.1 and 6561.3~\AA, showcasing the spectral profiles at different solar disk positions: the center, the east, and the west limbs.

A clear relative Doppler shift is evident between the solar \ion{Si}{1} line and the stationary telluric water lines, providing a direct spectroscopic signature of solar rotation. We quantified this kinematic pattern through double-Gaussian fitting. For the disk center and the west limb, where the lines were well-separated, the fit yielded a redshift of the \ion{Si}{1} line at the west limb relative to the disk center of approximately $1.894\ \mathrm{km\ s^{-1}}$. At the east limb, the \ion{Si}{1} line is severely blended with a telluric line due to its blueshift. By fixing the telluric line to its laboratory wavelength to constrain the fit, we resolved a \ion{Si}{1} blueshift of approximately $1.875\ \mathrm{km\ s^{-1}}$. The double-Gaussian fit also returned an average line broadening of approximately $13.3\ \mathrm{pm}$ for the \ion{Si}{1} line. After subtracting a linear continuum from the H$\alpha$ background and the telluric water line at 6560.50 Å, the corresponding fitted \ion{Si}{1} line profiles are overplotted as the solid curves in the bottom panel of Figure \ref{halpha}. This successful isolation and precise kinematic analysis of a weak photospheric line, amidst the strong H$\alpha$ chromospheric background and telluric contamination, unequivocally validates the exceptional spectral resolution, stability, and quantitative diagnostic capability of the VIPA spectrograph.
\section{Outlook}
\label{sec:discussion}

The successful on-sky validation of the VIPA prototype spectrograph confirms its design principle and core capabilities. To fully realize its potential for precision solar spectroscopy, the following areas are prioritized for further development:
\begin{itemize}[label={\fontsize{4}{4}\selectfont\textbullet}]

\item Simultaneous Spectroscopy for Multi-objects: 

a VIPA spectrograph can be used with a long entrance slit if the spectral range is within one free spectral range (FSR); in that case the spatial resolution along the slit is fully preserved. When the required spectral coverage spans several FSRs, an orthogonal (cross) disperser must be employed, which replaces the slit‑direction spatial information with the cross‑dispersion axis. Nevertheless, for multi‑object observations, a very practical alternative is to arrange several discrete targets (\eg , slits or fibres) along the cross‑dispersion direction. This allows simultaneous spectroscopy for multiple targets, provided that the required spectral range is not too wide and the detector is sufficiently large along the orthogonal direction. This approach is analogous to multi‑object spectroscopy with conventional grating spectrographs and is well suited for VIPA‑based instruments. 

    \item Advanced Instrumental Profile Characterization:
    
While analytical functions have been used to model the instrumental profile of our VIPA spectrograph, its inherently asymmetric, wavelength-dependent, and spatially variable nature poses a persistent challenge. Non-parametric methods, such as the Gaussian Process regression technique recently demonstrated by \citet{milakovic2024new}, offer a promising alternative for capturing these complex variations. The characteristics of VIPA line profiles align well with the challenges addressed in their work. Future efforts should therefore focus on adapting such a data-driven framework to the unique instrumental profile structure of VIPA spectrographs. This approach could provide a more flexible and accurate model than current analytical approximations, ultimately improving wavelength calibration and line center measurement precision.

    \item Optimal Frequency Comb Repetition Rate Configuration:
    
The current calibration accuracy is limited by the large comb line spacing ($\sim$16.6 FWHMs). 
The optimal repetition rate for the laser frequency comb must be specifically determined for VIPA spectrographs, as empirical rules from conventional grating instruments are not directly applicable. Dedicated simulations will be conducted to establish the relationship between the repetition rate and calibration accuracy in the VIPA context. Key parameters to consider include the VIPA's free spectral range, the width of its resolution element, and the necessary mode separation to balance calibration line density against blending. The outcome will be a set of VIPA-specific guidelines for selecting comb parameters to maximize calibration performance.

\item Laser Frequency Comb Simultaneous Calibration:

The current prototype lacks a dual-channel astrocomb for real-time calibration. For high-precision radial velocity and magnetic field measurements, however, simultaneous calibration is essential to correct for short-term instrumental drift. A dual-channel design—with one channel dedicated to the science target and the other to a stabilized frequency comb—would provide a continuous and precise wavelength reference. Although this configuration reduces the total spectral coverage per exposure, the impact can be mitigated by strategically observing only selected wavelength regions containing key diagnostic lines. Implementing such a system is a crucial step toward achieving the stability required for advanced solar and stellar spectroscopy.

    \item Rigorous Spectral Data Reduction Pipeline:
    
Our current pipeline requires significant enhancement to properly handle the complex instrumental profile of the VIPA spectrograph, which is crucial for achieving high-precision radial velocity and equivalent width measurements. Future development will focus on building a comprehensive framework based on forward modeling. This framework must accurately convolve the astrophysical spectrum with the asymmetric, spatially variable instrumental profile and robustly propagate uncertainties. Key implementations will include optimal extraction algorithms tailored to the VIPA's line spread function and a dedicated forward model capable of simultaneously fitting for astrophysical parameters and instrumental effects.
 
    \item Mechanical and Stability Enhancements:
    
The mechanical design of the current prototype can be significantly improved to enhance the stability and operational robustness. Key upgrades should include integrated thermal stabilization to minimize calibration drift, advanced vibration isolation to reduce mechanical noise, and precision alignment fixtures for the VIPA and optical train. Furthermore, compact and robust packaging will facilitate portability and field deployment. Implementing automated control and remote operation capabilities would also enable long-term monitoring campaigns. These refinements are essential to reduce systematic errors, unlock long-term stability, and broaden the practical applications of VIPA spectrographs.
   
\end{itemize}

Looking beyond the immediate improvements, the advanced VIPA spectrograph platform holds significant promise for future solar physics applications. Its combination of high spectral resolution and compact design is ideally suited for spectroscopic observations of the Sun, enabling simultaneous multi-target observations with high spectral-spatial resolution to study the complex solar atmospheric dynamics. In particular, the super-high spectral resolution observations allow for the detection of the slow rising motion of solar filaments at the very early stage of their eruption \citep{chen2011coronal,chen2020some}. Such detections of the very early precursors of filament eruptions are crucial in space weather forecast.
The integration with laser frequency combs will provide the absolute wavelength stability required for precise measurements of plasma velocities and magnetic fields. Furthermore, the intrinsically compact architecture makes it a compelling candidate for dedicated instruments on smaller solar telescopes or as a complementary module on larger facilities, and paves the way for miniaturized spectrographs on future space-based solar observatories.

\begin{acknowledgments}
The authors acknowledge support from the National Key Research and Development Program of China (Nos. 2024YFA1612002, 2024YFA1612003); the National Natural Science Foundation of China (Grant Nos. 11973009, 12293054, 12473087, 12127901, and 12325303); Jiangsu Provincial Key Research and Development Program (No. BE2023080); Chinese Academy of Sciences (No. KGFZD-145-23-04-03).
\end{acknowledgments}

\bibliographystyle{aasjournal}
\bibliography{SolarVbib}

@article{xiao2004dispersion,
  author={ Shijun {Xiao} and A. M. {Weiner} and C. {Lin}},
  journal={IEEE Journal of Quantum Electronics}, 
  title={A dispersion law for virtually imaged phased-array spectral dispersers based on paraxial wave theory}, 
  year={2004},
  volume={40},
  number={4},
  pages={420-426},}

@article{chen2020some,
  title={Some interesting topics provoked by the solar filament research in the past decade},
  author={Chen, Peng-Fei and Xu, Ao-Ao and Ding, Ming-De},
  journal={Research in Astronomy and Astrophysics},
  volume={20},
  number={10},
  pages={166},
  year={2020},
  publisher={IOP Publishing}
}

@article{chen2011coronal,
  title={Coronal mass ejections: models and their observational basis},
  author={Chen, Peng-Fei},
  journal={Living Reviews in Solar Physics},
  volume={8},
  number={1},
  pages={1--92},
  year={2011},
  publisher={Springer}
}

@article{hong2022statistical,
  title={Statistical analysis of the Si i 6560.58 {\AA} line observed by CHASE},
  author={Hong, Jie and Qiu, Ye and Hao, Qi and Xu, Zhi and Li, Chuan and Ding, Mingde and Fang, Cheng},
  journal={Astronomy \& Astrophysics},
  volume={668},
  pages={A9},
  year={2022},
  publisher={EDP Sciences}
}

@article{qiu2022calibration,
  title={Calibration procedures for the CHASE/HIS science data},
  author={Qiu, Ye and Rao, ShiHao and Li, Chuan and Fang, Cheng and Ding, MingDe and Li, Zhen and Ni, YiWei and Wang, WenBo and Hong, Jie and Hao, Qi and others},
  journal={Science China Physics, Mechanics \& Astronomy},
  volume={65},
  number={8},
  pages={289603},
  year={2022},
  publisher={Springer}
}

@article{li2022chinese,
  title={The Chinese H$\alpha$ solar explorer (CHASE) mission: An overview},
  author={Li, Chuan and Fang, Cheng and Li, Zhen and Ding, MingDe and Chen, PengFei and Qiu, Ye and You, Wei and Yuan, Yuan and An, MinJie and Tao, HongJiang and others},
  journal={Science China Physics, Mechanics \& Astronomy},
  volume={65},
  number={8},
  pages={289602},
  year={2022},
  publisher={Springer}
}

@article{radziemski1965arc,
  title={Arc spectrum of silicon},
  author={Radziemski Jr, Leon J and Andrew, Kenneth L},
  journal={Journal of the Optical Society of America},
  volume={55},
  number={5},
  pages={474--491},
  year={1965},
  publisher={Optical Society of America}
}

@article{zirker1993coronal,
  title={Coronal heating: Invited review paper},
  author={Zirker, Jack B},
  journal={Solar physics},
  volume={148},
  number={1},
  pages={43--60},
  year={1993},
  publisher={Springer}
}

@article{milakovic2024new,
  title={A new method for instrumental profile reconstruction of high-resolution spectrographs},
  author={Milakovi{\'c}, Dinko and Jethwa, Prashin},
  journal={Astronomy \& Astrophysics},
  volume={684},
  pages={A38},
  year={2024},
  publisher={EDP Sciences}
}

@article{unno1956line,
  title={Line formation of a normal Zeeman triplet},
  author={Unno, Wasaburo},
  journal={Publications of the Astronomical Society of Japan},
  volume={8},
  number={3-4},
  pages={108--125},
  year={1956},
  publisher={Oxford University Press}
}

@article{solanki2003sunspots,
  title={Sunspots: an overview},
  author={Solanki, Sami K},
  journal={The Astronomy and Astrophysics Review},
  volume={11},
  number={2},
  pages={153--286},
  year={2003},
  publisher={Springer}
}

@article{lites2005magnetic,
    author = {Lites, B. W.},
    title = {Magnetic Flux Ropes in the Solar Photosphere: The Vector Magnetic Field Under Active Region Filaments},
    journal = {The Astrophysical Journal},
    year = {2005},
    volume = {622},
    number = {2},
    pages = {1275},
    doi = {10.1086/428820}
}

@article{kuckein2012active,
    author = {Kuckein, C. and Martinez Pillet, V. and Centeno, R.},
    title = {An active region filament studied simultaneously in the chromosphere and photosphere. II. Doppler velocities},
    journal = {Astronomy \& Astrophysics},
    year = {2012},
    volume = {542},
    pages = {A112},
    doi = {10.1051/0004-6361/201218887}
}

@article{hao2018calibration,
  title={Calibration Tests of a 25-GHz Mode-spacing Broadband Astro-comb on the Fiber-fed High Resolution Spectrograph (HRS) of the Chinese 2.16-m Telescope},
  author={Hao, Zhibo and Ye, Huiqi and Han, Jian and Wu, Yuanjie and Zhai, Yang and Xiao, Dong},
  journal={\pasp},
  volume={130},
  number={994},
  pages={125001},
  year={2018},
  publisher={IOP Publishing}
}

@article{zhu2023dispersion,
  title={Dispersion characteristics of the multi-mode fiber-fed VIPA spectrograph},
  author={Zhu, Xiaoming and Lin, Dong and Zhang, Zhongnan and Xie, Xintong and He, Jinping},
  journal={The Astronomical Journal},
  volume={165},
  number={6},
  pages={228},
  year={2023},
  publisher={IOP Publishing}
}

@article{zhu2020vipa,
  title={A VIPA spectrograph with ultra-high resolution and wavelength calibration for astronomical applications},
  author={Zhu, Xiaoming and Lin, Dong and Hao, Zhibo and Wang, Liang and He, Jinping},
  journal={The Astronomical Journal},
  volume={160},
  number={3},
  pages={135},
  year={2020},
  publisher={IOP Publishing}
}

@article{bourdarot2017nano,
  title={NanoVipa: a miniaturized high-resolution echelle spectrometer, for the monitoring of young stars from a 6U Cubesat},
  author={Bourdarot, G and Le Coarer, E and Bonfils, X and Alecian, E and Rabou, P and Magnard, Y},
  journal={CEAS Space Journal},
  volume={9},
  number={4},
  pages={411--419},
  year={2017},
  publisher={Springer}
}

@article{shirasaki1996large,
  title={Large angular dispersion by a virtually imaged phased array and its application to a wavelength demultiplexer},
  author={Shirasaki, M},
  journal={OptL},
  volume={21},
  number={5},
  pages={366--368},
  year={1996},
  publisher={Optical Society of America}
}

@inproceedings{conconi2013espresso,
  title={ESPRESSO APSU: simplify the life of pupil slicing},
  author={Conconi, P and Riva, M and Pepe, F and Zerbi, FM and Cabral, A and Cristiani, S and Megevand, D and Landoni, M and Span{\'o}, P},
  booktitle={Novel Optical Systems Design and Optimization XVI},
  volume={8842},
  pages={178--186},
  year={2013},
  organization={SPIE}
}

@article{pepe2000harps,
  title={HARPS: a new high-resolution spectrograph for the search of extrasolar planets},
  author={Pepe, Francesco and Mayor, Michel and Delabre, Bernard and Kohler, Dominique and Lacroix, Daniel and Queloz, Didier and Udry, Stephane and Benz, Willy and Bertaux, Jean-Loup and Sivan, Jean-Pierre},
  journal={\procspie},
  volume={4008},
  pages={582--592},
  year={2000},
  publisher={International Society for Optics and Photonics}
}

@article{smitha2023non,
  title={Non-LTE formation of the Fe I 6173 {\AA} line in the solar atmosphere},
  author={Smitha, HN and van Noort, M and Solanki, SK and Dur{\'a}n, JS Castellanos},
  journal={Astronomy \& Astrophysics},
  volume={669},
  pages={A144},
  year={2023},
  publisher={EDP Sciences}
  }

@article{wang2021high,
  title={High-resolution He I 10830 {\AA} narrowband imaging for a small-scale chromospheric jet},
  author={Wang, Ya and Zhang, Qingmin and Ji, Haisheng},
  journal={The Astrophysical Journal},
  volume={913},
  number={1},
  pages={59},
  year={2021},
  publisher={IOP Publishing}
}

@article{du2008properties,
  title={Properties of the He I 10830 {\AA} Line in Solar Flares},
  author={Du, Qiu-Sheng and Li, Hui},
  journal={Chinese Journal of Astronomy and Astrophysics},
  volume={8},
  number={6},
  pages={723},
  year={2008},
  publisher={IOP Publishing}
}

@article{lohner2018convective,
  title={Convective blueshifts in the solar atmosphere-I. absolute measurements with LARS of the spectral lines at 6302 {\AA}},
  author={L{\"o}hner-B{\"o}ttcher, Johannes and Schmidt, Wolfgang and Stief, Franziska and Steinmetz, Tilo and Holzwarth, Ronald},
  journal={Astronomy \& Astrophysics},
  volume={611},
  pages={A4},
  year={2018},
  publisher={EDP Sciences}
}

@article{lemke2016gottingen,
  title={The G{\"o}ttingen Solar Radial Velocity Project: Sub-m s- 1 Doppler Precision from FTS Observations of the Sun as a Star},
  author={Lemke, Ulrike and Reiners, Ansgar},
  journal={Publications of the Astronomical Society of the Pacific},
  volume={128},
  number={967},
  pages={095002},
  year={2016},
  publisher={IOP Publishing}
}

@article{de2007chromospheric,
  title={Chromospheric Alfv{\'e}nic waves strong enough to power the solar wind},
  author={De Pontieu, B and McIntosh, SW and Carlsson, M and Hansteen, VH and Tarbell, TD and Schrijver, CJ and Title, AM and Shine, RA and Tsuneta, S and Katsukawa, Y and others},
  journal={science},
  volume={318},
  number={5856},
  pages={1574--1577},
  year={2007},
  publisher={American Association for the Advancement of Science}
}

@article{stenflo2015fts,
  title={FTS atlas of the Sun’s spectrally resolved center-to-limb variation},
  author={Stenflo, Jan Olof},
  journal={Astronomy \& Astrophysics},
  volume={573},
  pages={A74},
  year={2015},
  publisher={EDP Sciences}
}

@article{doerr2012laser,
  title={A laser frequency comb system for absolute calibration of the VTT echelle spectrograph},
  author={Doerr, H-P and Steinmetz, Tilo and Holzwarth, Ronald and Kentischer, T and Schmidt, W},
  journal={Solar Physics},
  volume={280},
  number={2},
  pages={663--670},
  year={2012},
  publisher={Springer}
}

@article{liu2013horizontal,
  title={Horizontal flows in the photosphere and subphotosphere of two active regions},
  author={Liu, Yang and Zhao, Junwei and Schuck, PW},
  journal={Solar Physics},
  volume={287},
  number={1},
  pages={279--291},
  year={2013},
  publisher={Springer}
}



\end{document}